\documentclass[twocolumn,aps,superscriptaddress,showpacs,prl]{revtex4}
\usepackage{amsfonts,amssymb}
\usepackage{graphicx,latexsym,endnotes}
\usepackage{psfrag,amsmath}
\usepackage{pstricks}
\usepackage[english]{babel}

\begin{document}

\author{M. Zamparo} \email{marco.zamparo@pd.infn.it}
\affiliation{Dipartimento di Fisica G. Galilei and CNISM, Universit\`a
  di Padova, v. Marzolo 8, PD--35131 Padova, Italy}

\author{A. Trovato} \email{antonio.trovato@pd.infn.it}
\affiliation{Dipartimento di Fisica G. Galilei and CNISM, Universit\`a
  di Padova, v. Marzolo 8, PD--35131 Padova, Italy}

\author{A. Maritan} \email{amos.maritan@pd.infn.it}
\affiliation{Dipartimento di Fisica G. Galilei and CNISM, Universit\`a
  di Padova, v. Marzolo 8, PD--35131 Padova, Italy}
\affiliation{INFN, Sezione di Padova, PD--35131 Padova, Italy}

\title{A simplified exactly solvable model for $\beta$-amyloid aggregation}

\begin{abstract}
We propose an exactly solvable simplified statistical mechanical model
for the thermodynamics of $\beta$-amyloid aggregation, generalizing a
well--studied model for protein folding. The monomer concentration is
explicitly taken into account as well as a non trivial dependence on
the microscopic degrees of freedom of the single peptide chain, both
in the $\alpha$-helix folded isolated state and in the fibrillar
one. The phase diagram of the model is studied and compared to the
outcome of fibril formation experiments which is qualitatively
reproduced.
\end{abstract}

\maketitle

Amyloids are insoluble fibrillar aggregates of proteins, stabilized
mostly by hydrogen bonds and hydrophobic interactions. They are
implicated in debilitating human pathologies, such as Alzheimer's,
Parkinson's disease and spongiform encephalopathies. Citotoxic species
have been recently identified with transient soluble oligomeric
structures whereas amyloid fibrils are believed to be the final most
stable state of the aggregation process \cite{ChitiDobson}. Virtually
all proteins can be induced to adopt the amyloid structure upon
appropriate conditions \cite{Fandrich}.

\begin{figure}
\center 
\includegraphics[width=6cm]{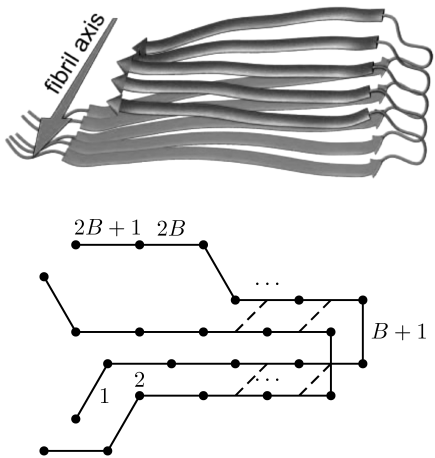}
\caption{In--register parallel $\beta$-hairpins in A$\beta40$
peptide structural model \cite{Petkova}(top) and schematic
representation of two interacting monomers in our model (bottom).
Dots correspond to aminoacids, horizontal and vertical segments are
ordered peptide bonds and tilted segments are unstructured
ones. Dashed lines represent contacts between the two monomers. See
text for details.}
\label{fig:intro}
\end{figure}

A common signature of fibril formation is the presence of a stable
core of cross-$\beta$ structure, with $\beta$-strands running
orthogonal to the fibril axis and forming several $\beta$-sheets which
may intertwine along the latter. The cross-$\beta$ structure is
identified through its typical X-rays diffraction pattern and binding
to specific fluorescent dyes. More sophisticated techniques, such as
solid state NMR, are needed in order to provide structural models at
atomic level. In a few known cases, for intermediate chain lengths in
between 20 and 40, all peptide monomers may adopt a repeating hairpin
structure within the fibrillar aggregate \cite{Petkova,Ferguson}.
This is then stabilized by interchain hydrogen bonds between the same
residues in different chains, leading to the so--called parallel
in--register arrangement \cite{Plos06,Eisenberg} shown in
Fig.\ \ref{fig:intro}.

The conformational ensembles populated at low concentration by
proteins, which aggregate into amyloid fibrils at higher density, may
vary from the large amount of fluctuating structures of natively
unfolded proteins and peptides, such as the A$\beta$-peptide related
to Alzheimer's, to the well defined structures of globular
proteins. In the latter case, the competition between the stability of
the native structure and of the amyloid fibrils is crucial in
determining the amyloidogenic behavior \cite{ChitiDobson2}.

In the context of protein folding, simple models based on the geometry
of the native structure have been very useful in unraveling folding
kinetics. In the same manner, one can speculate that the geometry of
the fibrillar aggregate, as typified by the parallel in--register
hairpin structure, may play a similar role in aggregation
kinetics. Within this spirit, the competition described above for the
aggregation of globular proteins becomes a competition between two
alternative geometries, which needs to be assessed already at
equilibrium.

The purpose of the present Letter is proposing a simplified
statistical mechanical model for $\beta$-amyloid aggregation,
generalizing a well--studied one for protein folding. Our model
explicitly depends on protein concentration and has the virtue of
being exactly solvable. For more realistic descriptions, even at a
coarse--grained level, the computational cost of achieving
thermodynamic equilibrium at different concentrations is
prohibitive. On the other hand, here we consider a non trivial
dependence on the microscopic degrees of freedom of the single peptide
chain, both in the folded and in the fibrillar state. Other simplified
models describe monomers through just a few macrostates
\cite{Ferrando,Nicodemi,Lee}. Notably, we succeed in reproducing, at
least qualitatively, the behavior of fibril formation experiments in
the presence of the denaturant trifluoroethanol (TFE) in different
concentrations.

Our model starts from the one introduced by Wako and Sait\^{o}
\cite{WS1,WS2} and then reconsidered by Mu\~{n}oz, Eaton and
co--workers \cite{ME1,ME2,ME3} (WSME--model). The latter has been the
subject of many works with applications to real proteins
\cite{BP,ItohSasai1,AbeWako,ZP1,BPZ2,IPZ1,IP}.  Despite its
simplicity, it has been able to capture the main features of the
kinetic behavior and folding pathways of specific molecules.

The WSME--model is a highly simplified model of the protein folding
process built on the premise that the latter is mainly determined by
the structure of the native functional state, whose knowledge is
assumed. Only native interactions are included, classifying the model
as G\={o}--like \cite{Go}. Moreover, the interaction between two
aminoacids in the protein sequence is possible only if all intervening
peptide bonds are in their native conformation. The entropy loss due
to fixing peptide units in this conformation is finally explicitly
taken into account.

Within this framework, a polypeptide chain made up of~$N+1$ aminoacids
is described as a sequence of $N$ peptide bonds. Two conformations are
considered for each bond: the native one and a generic disorder state.
Thus, a binary variable $m_i$ is associated to the $i$-th peptide
unit, taking value 1 and 0 in the two cases respectively, and the free
energy $F$ of the model can be written in unit of $k_BT$, with $T$ the
absolute temperature, as
\begin{equation}
F(m)=\sum_{i=1}^{N-1}\sum_{j=i+1}^N\epsilon_{ij}\Delta_{ij}\prod_{k=i}^{j}m_k 
-\sum_{i=1}^Nq_i(1-m_i).
\label{def:WSME}
\end{equation}
The contact matrix, with entry $\Delta_{ij}$ equal to 1 if the
$i$-th\\ and $j$-th bonds are close to each other in the native
structure and equal to 0 otherwise, tell us which are the native
interactions.  Their energetic amount is then quantified by the
dimensionless contact energy $\epsilon_{ij}<0$, referring to the
$i$-th and $j$-th peptide units.  This contributes to the free energy
only if the product $\prod_{k=i}^{j}m_k$ does not vanish, that is only
if such two bonds are the ends of a sequence of ordered peptide units,
thus realizing the depicted interaction. Finally, recognizing the
microscopic multiplicity of an abstract disorder state, an entropic
cost $q_i>0$ is given to the ordering of the $i$-th peptide bond.
 
Our model is an extension of the WSME--model, suitable for the
thermodynamics of $\beta$-amyloid aggregation. The basic idea is that
peptide monomers can either fold into their native structure or
partially lose this feature before aggregating in fibrils.  Here we
focus, for simplicity, on $\alpha$-helices while the aggregation is
assumed to require a hairpin shape and proceed by parallel
in--register arrangement as in Fig.\ \ref{fig:intro}, thus mimicking
real fibrils.  Other in--register amyloid structures with more than
two $\beta$-layers \cite{Kajava2004} could be also implemented.

We will define the model in a bottom-up approach.  Let us begin
introducing the free energy of isolated monomers, which can fold into
$\alpha$-helix native structure. In such a structure an hydrogen bond
is formed between peptide units $i$ and $j$ so that $|i-j|=3$. Then,
for a homogeneous molecule with an odd number of peptide bonds,
$2B+1$, following Eq.\ (\ref{def:WSME}) we choose the free energy as
\begin{equation}
F(m,0)=-\epsilon_{\alpha}\sum_{i=1}^{2B-2}\prod_{k=i}^{i+3}m_{k}-q\sum_{i=1}^{2B+1}(1-m_{i}),
\label{def:alpha}
\end{equation} 
because $\Delta_{ij}=1$ if $|i-j|=3$ and 0 otherwise. The
dimensionless parameters $\epsilon_{\alpha}>0$ and $q>0$ account
respectively for the energy strength of each contact and the entropic
cost of ordering each bond.

As far as the interaction between different peptides is concerned, we
assume that aggregation involves and requires a partial
$\beta$-hairpin shape, which is obtained by removing some helical
contacts. In such a view, the small loop region of the hairpin formed
by a monomer with $2B+1$ peptide units is identified with the peptide
bond $B+1$, from which two strands depart as shown in
Fig.\ \ref{fig:intro}.\\ Fibril formation is triggered by pairing a
part of the ordered fragments of the two strands from one molecule
with the same part of another. A measure of the ``$\beta$-order''
extent associated to a pair of consecutive $\beta$-hairpins, with
WSME--variables $m$ and $m'$, is provided by
$\mathcal{B}(m,m')=\sum_{i=1}^{B}\prod_{k=i}^{2B+2-i}m_{k}m'_{k}$ and
vanishes if loop regions are not both ordered. Otherwise, it is the
common number of ordered peptide units facing each other beginning
from loops. For the case shown in Fig.\ \ref{fig:intro}, we have
$\mathcal{B}(m,m')=2$.

We can then interpret the aggregation phenomenon, which is driven and
stabilized by hydrogen bonds, as the formation of $2c$ contacts
between the $\beta$-portions of the two different monomers, where $c$
has thus to be in between 0 and $\mathcal{B}(m,m')$. We assume that
the pairing between different peptides starts from loops and go on
sequentially along the strands, suggesting the idea that these
regions, having the same shape, are the most suitable to initiate the
aggregation. In equilibrium conditions this mechanism corresponds to
assume that there is only one way to form the above $2c$ contacts. In
Fig.\ \ref{fig:intro} all available interactions of this kind are
present.

A segment can gain energy being either in a helical state and
unbounded by other peptides or in a $\beta$-hairpin state and bounded
to another hairpin. We assume that, if a molecule binds another one
with $2c>0$ hydrogen bonds, then the helical contacts including
peptide bonds participating to the pairing, that is in the stretch
going from $B-c+1$ to $B+c+1$, are suppressed. Hence, the free energy
of that monomer becomes
\begin{equation}
F(m,c)=F(m,0)+\epsilon_{\alpha}\sum_{i=\max\{B-c-2,1\}}^{\min\{B+c+1,2B-2\}}
\prod_{k=i}^{i+3}m_{k},
\end{equation}  
being $F(m,0)$ the free energy of the isolated $\alpha$-helix defined
by Eq.\ (\ref{def:alpha}). In turn, we shall denote by $\epsilon_{\beta}>0$
the energetic gain, in unit of $k_BT$, of one contact between
different monomers.

Now we take into account the translational and rotational entropy loss
$S(c)\ge 0$ due to the aggregation of different peptides with the
formation of $2c$ hydrogen bonds. We will choose $S$ so that
$S(0)=0$. At last, the free energy for a system of two close molecules
that can aggregate takes the form
$F(m,c)+F(m',c)-2\epsilon_{\beta}c+S(c)$, with the constraint
$c\le\mathcal{B}(m,m')$.

We want to stress that the $\beta$-hairpin shape is not needed {\it a
  priori} in order to have aggregation between different monomers, but
it is rather considered as a concomitant event to the matching
process. Moreover, the requirement of ordered stretches of peptide
units to form contacts between molecules is just a way to express that
only few chain conformations are suitable for
aggregation. Intra--helix contacts represent general native
interactions protecting isolated conformers from the aggregation--prone
states \cite{ChitiDobson2}.

\begin{figure}
\center
\includegraphics[width=8.5cm]{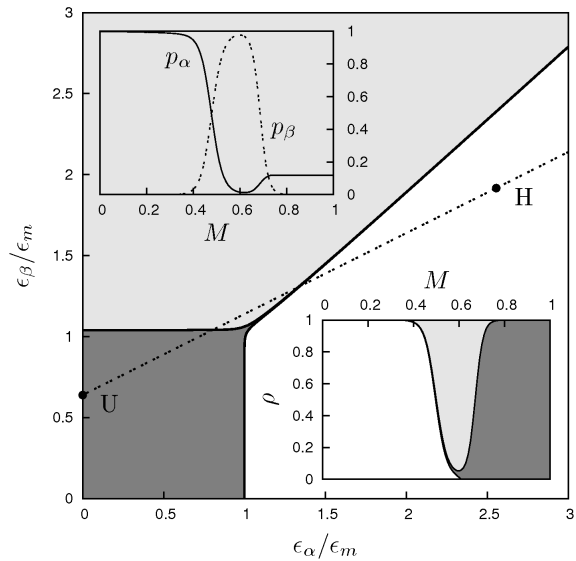}
\caption{Phase diagram of the model in the plane
  $(\epsilon_\alpha,\epsilon_\beta)$ at $\rho=0.7$ and $\sigma=5$,
  being $\sigma$ the entropy loss associated to the aggregation. Here
  $B=10$ corresponding to peptides of 21 monomers. White, light--gray
  and gray regions are helix, fibril and unfolded phases respectively.
  Top inset depicts the order parameters $p_{\alpha}$ and $p_{\beta}$
  as a function of the TFE concentration $M$ while the bottom one
  shows the phase diagram in the plane $(M,\rho)$. See text for
  details.}
\label{fig:phases}
\end{figure}

Finally, we model the formation of an aggregate as a growth of a
``one--dimensional structure''. To this aim, we describe a system of
many peptides by placing them on distinct sites $l$, $l=1,2,\ldots,L$,
of a one--dimensional lattice and including in the model only
interactions between nearest--neighbor molecules. The occupation
number $n_l$ of site $l$ is 1 if a monomer is present in that position
and 0 otherwise.  Furthermore, to each site we associate
WSME--variables describing the conformation of the peptide chain
placed there and thus $m_{l,i}$ will give the state of the $i$-th
peptide unit of the molecule at $l$.  In order to avoid an unphysical
entropic contribution, we set $m_{l,i}$ at 0 for any $i$ if
$n_l=0$. For simplicity, the symbol $m_l$ will be used for the array
$(m_{l,1},m_{l,2},\ldots,m_{l,2B+1})$ of binary variables related to
the node $l$, $m$ for all these variables and $n$ for the collection
of occupation numbers.  Finally, we need the variable $c_l$ keeping
count of the contacts between close molecules residing at nodes $l$
and $l+1$. This variable ranges from 0 to $\mathcal{B}(m_l,m_{l+1})$
and, as expected, no interaction is possible between sites $l$ and
$l+1$ when they are not both occupied.

The free energy $H_L$ of the full model is then a generalization of
the one introduced above for two monomers. Using the dummy variables
$c_0=0$ and $c_L=0$ and noticing that the number of peptide units of
the molecule at site $l$ involved in contacts with other molecules is
properly related to $\max\{c_{l-1},c_l\}$, this free energy reads
\begin{eqnarray}
\nonumber
H_L(n,m,c)&=&\sum_{l=1}^Ln_lF(m_l,\max\{c_{l-1},c_l\})\\ 
&-&\sum_{l=1}^{L-1}\biggl[2\epsilon_{\beta}c_l-S(c_l)\biggr]
-\mu\sum_{l=1}^Ln_l.
\label{def:H}
\end{eqnarray}
The contribution of the chemical potential $\mu$, which will be
determined by imposing a given value $\rho$ to the monomer density,
has been here included. 

The Boltzmann distribution with the free energy of Eq.\ (\ref{def:H})
provides the possibility to evaluate equilibrium expectation values of
physical observables. The present model can be solved exactly by means
of a transfer matrix method, because of the presence of short--range
interactions in a 1-dimensional system and the possibility of exactly
tracing on the WSME--variables.  Details are shown in the
supplementary material \cite{EPAPS}. Here we restrict to some results
on the behavior of two order parameters related to the fraction of
isolated helices and aggregated molecules. The former, $p_{\alpha}$
\cite{EPAPS}, measures the global order of peptides when they do not
interact at all and is defined as the equilibrium average of the
fraction of native bonds per site, normalized to the density $\rho$,
considering only microscopic configurations which exclude aggregation
phenomena. The latter, $p_{\beta}$ \cite{EPAPS}, accounts for bonds
between different monomers and is given by the fraction of formed
contacts between two consecutive lattice nodes, again normalized with
respect to $\rho$.

Fig.\ \ref{fig:phases} shows the phase diagram of the model in the
thermodynamic limit $L\to\infty$, where different phases are separated
by the conditions $p_{\alpha}=1/2$ and $p_{\beta}=1/2$.  Here we
choose $B=10$ and $q=2$ but other values of $q$ only marginally affect
this diagram. Moreover, we consider the case $S(0)=0$ and
$S(c)=\sigma>0$ independent of $c$ if $c=1,2,\ldots,B$, assuming that
most of the entropy loss in aggregation is due to the formation of
just one contact between monomers. Parameters $\epsilon_\alpha$ and
$\epsilon_\beta$ are referred to the midpoint $\epsilon_m\sim 2.35$,
depending on both $B$ and $q$, of an helix in the pure
WSME--model. The energy scale $\epsilon_m$ may be obtained by imposing
$p_\alpha=1/2$ when the density $\rho$ approaches 0.

Three regions are recognized. The first is the region of unfolded
isolated peptides where both $p_{\alpha}$ and $p_{\beta}$ are less
than 1/2. For small $\epsilon_\alpha$ the order parameter $p_{\alpha}$
is closed to the value $1/(1+\text{e}^q)\sim 0.12$ of a completely
unfolded structure in the WSME--model.  The second region, where
$p_{\alpha}>1/2$ and $p_{\beta}<1/2$, corresponds to native isolated
helical peptides. The boundary between the unfolded region and the
$\alpha$-helix region is weakly depending on both $\sigma$ and $\rho$
and almost coincides with the one obtainable in the plain WSME--model
for the same helices.  Finally, there is the fibril region where
$p_{\alpha}<1/2$ and $p_{\beta}>1/2$. Increasing $\rho$ or decreasing
$\sigma$ favors the aggregation by lowering the boundary between this
region and the denatured one.

Since the energetic parameters $\epsilon_\alpha$ and $\epsilon_\beta$
are effective parameters mediated by solvent, we may expect them to
vary in a non trivial way as external conditions, such as temperature,
different denaturant concentrations, solution ionic strength and pH,
are changed. The denaturant agent TFE is commonly used in fibril
formation assays because, at moderate concentrations, it disrupts the
native structures of isolated proteins, without preventing the
formation of inter--molecular contacts \cite{TFE}. At high
concentrations, TFE addition results in the stabilization of isolated
unfolded proteins \cite{TFE}.

We can mimic the TFE effect by assuming that both $\epsilon_\alpha$
and $\epsilon_\beta$ are simple linear decreasing functions of its
concentration $M$, with $\epsilon_\alpha$ decreasing more than
$\epsilon_\beta$. For example, by moving along the straight line in
Fig.\ \ref{fig:phases} from H at $M=0$ to U at $M=1$, the observed
native--fibril--unfolded pattern \cite{TFE} can be qualitatively
reproduced. Given such a dependence of $\epsilon_\alpha$ and
$\epsilon_\beta$ on $M$, the top inset in Fig.\ \ref{fig:phases}
depicts the profile of $p_\alpha$ and $p_\beta$ as a function of the
TFE concentration whereas the bottom one reports the phase diagram of
the model in the plane $(M,\rho)$. The pattern discussed above is
present for high values of peptide density, with the fibril stability
interval in TFE concentration narrowing with decreasing peptide
density. At low density the fibril phase is not present anymore and
the peptides remain always isolated going directly from the native to
the unfolded state, with increasing TFE concentration.

In summary, in this Letter we have proposed a highly simplified
equilibrium model to describe the aggregation of identical monomers
and the consequent formation of fibrillar structures. Despite its
simplicity, the model has been shown to explain different phases of
the system, such as unfolded and aggregated states, and to
reproduce qualitatively the observed trend of fibril formation
experiment as a function of trifluoroethanol concentration. Moreover,
we argue that a kinetic version of the model could shed new light on
the protein aggregation dynamics and work is in progress along this
line.

This work has been supported by the Italian Ministry of Education,
University and Research via the PRIN 2007B57EAB and by University of
Padua via Progetto di Ateneo CPDA083702.

\end{document}